%% file: main.tex
\documentclass[12pt]{article}

\usepackage[utf8]{inputenc}
\usepackage{charter}
\usepackage{amsmath} 
\usepackage{amssymb}
\usepackage{fullpage}
\usepackage{hyperref}
\usepackage{graphicx}
\usepackage{lipsum}
\usepackage[sort&compress,numbers]{natbib}
\usepackage{hyperref}
\hypersetup{hidelinks}
\usepackage[total={6.5in,9in},top=1in,headsep=0.1in,headheight=1in]{geometry}
\usepackage[dvipsnames,table,xcdraw]{xcolor}

\newcommand\snowmass{
\begin{center}
  \rule[-0.2in]{\hsize}{0.01in}\\
  \rule{\hsize}{0.01in}\\
  \vskip 0.1in
  Submitted to the Proceedings of the US Community Study\\ 
  on the Future of Particle Physics (Snowmass 2021)\\
  \rule{\hsize}{0.01in}\\
  \rule[+0.2in]{\hsize}{0.01in}\\[-2em]
\end{center}
}

\usepackage[firstpage=true]{background}
\backgroundsetup{contents={\parbox{6.5in}{\snowmass}}, scale=1,placement=top,opacity=1,color=black,position={3.25in,1.2in}}

\usepackage{fancyhdr}
\fancypagestyle{plain}{%
  \fancyhf{}%
  \fancyhead[C]{}
  \fancyfoot[C]{\thepage}
}

\fancypagestyle{empty}{%
  \fancyhf{}%
  \fancyhead[C]{{\it Snowmass2021 Cosmic Frontier White Paper: Dark Matter Searches with DESI}}
  \fancyfoot[C]{\thepage}
}
\input{newcommands.tex}
\title{Snowmass2021 Cosmic Frontier White Paper: \\ Prospects for obtaining Dark Matter Constraints with DESI}

\input{authors.tex}

\begin{document}

\maketitle
\begin{center}
\centering{\bf \Large Executive Summary}
\end{center}

For the past several decades efforts to directly detect dark matter (DM), the particle thought to constitute $\sim 24\%$ of the matter--energy density of the universe have largely focused on cold dark matter. Despite heroic and sustained efforts, highly sensitive direct-detection experiments have not yet led to the discovery of the WIMP (although the allowed mass/energy scale and interaction cross-section have been significantly constrained; see, e.g., \cite{Zyla:2020zbs} for a recent review). This has led to increasing interest in alternatives to
the $\Lambda$CDM paradigm and a growing impetus to explore
alternative DM scenarios (including fuzzy DM, warm DM, self-interacting DM, etc.). In many of these scenarios, particles cannot be detected directly and constraints on the properties of these DM particles can ONLY be arrived at using astrophysical observations. 

The Dark Energy Spectroscopic Instrument (DESI), which recently began its regular operations, is currently one of the most powerful instruments for wide-field surveys.
The synergy of DESI with current (e.g. ESA's \Gaia \ satellite) and future observing facilities including the James Webb Space Telescope (JWST), the Vera Rubin Observatory's Legacy Survey of Space and Time (LSST), and the Nancy Grace Roman Space Telescope's High Latitude Survey (HLS) will yield datasets of unprecedented size and coverage that will enable constraints on DM over a wide range of physical and mass scales and across redshifts. 

DESI will obtain spectra of the Lyman-$\alpha$ (Ly-$\alpha$) forest out to $z\simeq 5$ and radial velocities of $\simeq$~10 million stars in the Milky Way (MW) and Local Group satellites.
DESI will detect four times as many Ly-$\alpha$ QSOs than were observed in the largest previous survey, yielding about 1 million spectra, which will put constraints on clustering of the low-density intergalactic gas.  Radial velocities from DESI in conjunction with astrometry from \Gaia~(and eventually Roman's HLS) will enable us to constrain the global distribution of DM within the MW and in dwarf satellites, and the mass function of dark subhalos (the DM distribution on small scales). This paper describes how DESI will enable tighter constraints on the nature of DM if the survey is supported by a large simulation and modeling effort.

The paradigm that the formation of cosmological structure can be described as a solution to Boltzmann equations has been extensively tested with simulations in many different environments --- from high redshifts and using the gas between galaxies to low redshifts and dwarf galaxies, all the way to kinematics within our Galaxy. The majority of these simulations to date have focused on collisionless CDM. Simulations with alternatives to CDM have recently been gaining ground but are still in their infancy.  Specifically, while there are numerous publicly available large-box and zoom-in simulations in the $\Lambda$CDM framework, there are no comparable publicly available WDM, SIDM, FDM simulations. DOE should fund such a public simulation suite to enable a more cohesive community effort to compare observations from DESI (and other surveys) with numerical predictions. To aid such comparisons it is also necessary to support the creation of suites of queryable mock datasets from simulations which account for selection effects and survey volume. Machine learning (deep learning) tools should be supported with the goal of assessing (from blinded mock datasets) the likelihood that a given set of observational constraints on DM is consistent with simulation predictions within a specific DM scenario. Such an effort will greatly enhance the impact of DESI on DM science.

\section{Introduction}

Much of what we know about dark matter (DM) on galactic scales comes from comparing the predictions of cosmological simulations with  the observed properties of galaxies on these scales. Cosmological simulations make several key predictions about the DM halos of galaxies that can be validated using the phase-space coordinates of stars in the Milky Way and in Local Group dwarf galaxies as well as with the structure of the Lyman-$\alpha$ forest at high redshifts. These predictions differ for different types of DM in ways that reflect both the effect of the DM particle on primordial structure formation as well as possible interactions between the DM particles themselves and interactions between the DM particles and baryonic matter.

The design of DESI was optimized for cosmological galaxy redshift surveys, since the primary objective of the DESI survey is to probe the origin of cosmic acceleration by employing the baryon acoustic oscillation technique, with redshifts out to $z=3.7$. However, DESI will also obtain spectra of the Ly-$\alpha$ forest out to $z\approx5$ using a sample of QSOs $\sim 4$ times as large as the previous comparable spectroscopic survey. During bright time (unsuitable for cosmological targets), the project will measure the radial velocities of $\approx 10$ million stars in the Milky Way (MW) and in Local Group dwarfs. DESI data will enable models of the DM distribution both at high redshifts that contribute to the Ly-$\alpha$ forest and in the Milky Way and its vicinity.

\paragraph{The DESI Instrument and Survey:}
The Dark Energy Spectroscopic Instrument (DESI) is currently the premier multi-object spectrograph for wide-field surveys \citep[][]{desiScience,desiInstrument}. DESI deploys 5000 fibers over a $3.1^\circ$ diameter field of view at the prime focus of the Mayall 4m telescope at Kitt Peak National Observatory. The fibers feed ten identical 3-arm spectrographs, each spanning 3600-9824 \AA\ at a FWHM resolution of about 1.8 \AA.
Each fiber can be positioned individually by a robotic actuator within a radius of $1.48$ arcmin, with a small overlap between the patrol regions of adjacent fibers. The full set of fiber positioners can be reconfigured in 3-5 minutes. This design allows large areas of sky to be surveyed very rapidly at a density of $\sim 600$ targets per square degree. 

\paragraph{The DESI Milky Way Survey:}
The DESI Milky Way Survey (MWS) will be carried out alongside its primary 5-year cosmological program.  The main DESI-MWS is leveraging the already considerable impact of the European Space Agency's {\Gaia} mission by targeting stars between $16 < G \leq 20.5$. {\Gaia\,}  has revolutionized our understanding of the Milky Way by delivering precision astrometric parallaxes, proper motions and photometry for more than a billion stars to $G \simeq  20-21$~mag. {\Gaia’s} proper motion accuracy is $\simeq 0.25$~ mas/yr at the faint limit, corresponding to a velocity accuracy of tens of km/s at characteristic halo distances of 50 kpc.  The {\Gaia\,} Radial Velocity Spectrograph will also obtain radial velocities for the brighter stars ($\sim 33$ million stars  with $G<14$ in \Gaia\, DR3).
In contrast DESI-MWS will obtain radial velocities for stars down to $r=19.5$ (which is approximately $G=19.5$ for redder stars) and with velocity errors of $\lesssim 10\kms$. DESI which has outstanding survey speed due to its sensitive spectrographs and robotically controlled fiber positioners, is able to achieve this radial velocity accuracy with $\sim 300$ seconds of effective exposure time or less.

The MWS is primarily aimed at obtaining kinematics of the stellar halo and thick disk and will also study the stellar populations of the Milky Way by yielding stellar parameters ($T_{\rm eff}$, $\log{g}$, [Fe/H], [$\alpha/$Fe], and  elemental abundances for a few additional elements). MWS will have greater depth, greater homogeneity, and will result in an order-of-magnitude increase in sample size compared to existing stellar surveys with similar spectral resolution. The main MWS sample will focus on large-scale spatial and kinematic structure up to 100 kpc from the Sun. Radial velocities from MWS combined with {\Gaia\,} astrometry will constrain the 3D distribution of dark matter in the Galaxy. Radial velocities of stars in thin, dynamically cold stellar tidal streams will enable us to set constraints on the population of dark subhalos in the Galaxy that could also be used as a powerful tool to discriminate between different cosmological scenarios. DESI's large field of view makes it an efficient instrument for obtaining radial velocities of stars in dwarf satellites in the Local Group, especially in the outer parts of these galaxies where evidence of tidal disruption is easiest to observe. Such radial velocities, in combination with proper motions (0.25$\mu$as~yr$^{-1}$) from the Roman Space telescope will enable the tightest constraints to date on the matter density profiles of dwarf galaxies. This, in turn, will  allow us to obtain more accurate predictions of dark matter-induced photon fluxes from these objects. Such predictions are critical to setting more reliable and robust constraints on the properties of the dark matter particle. Finally, the DESI Backup Program (a program which will operate during very bright and poor seeing conditions when the main program cannot be successfully executed) will obtain spectra of $\simeq 8 $~million stars including those closer to the MW disk plane enabling constraints on the local DM density - important for direct DM detection experiments.

\paragraph{The DESI Ly-$\alpha$ Survey:}\label{sec:Lya-survey}

The Ly-$\alpha$ forest is detected as the series of absorption lines in the quasar spectra, caused by the Lyman-alpha transitions of neutral hydrogen in the low-density, high-redshift intergalactic medium (IGM). It is a biased continuous tracer of the quasi-linear matter density field, and the auto (cross) correlation function of the forests (with quasars) have been used to detect the Baryon Acoustic Osillations (BAO) signal\citep{dr16completed}. Furthermore, dark matter models can be tested using the 1D power spectrum $P_{\rm 1d}(k)$ of the Ly-$\alpha$ forest -- measuring the clustering down to small ($\sim$~Mpc) scales along the radial direction while averaging over all lines-of-sight -- allowing constraints to be placed on the underlying dark matter distributions from large scales to small scales.

DESI is designed to obtain about a million Ly-$\alpha$ quasars, as compared to the $\sim 200,000$ Ly-$\alpha$ quasar data set of the Sloan Digital Sky Survey's (SDSS) extended Baryon Oscillation Spectroscopic Survey (BOSS)\citep{dr16completed}, the largest existing data set available.
The observations are conducted with a 310\,deg$^{-2}$ target density, and the combination of the larger mirror and improved target selection allow a fainter limiting magnitude of $\textit{r}\sim23$ compared to previous experiments. This enables DESI to detect quasars with a target density of 200\,deg$^{-2}$, of which more than 60 quasars deg$^{-2}$ probe $z>2.1$ and thus contain an observable part of the Ly-$\alpha$ forest. It also allows one to probe higher redshifts $z>4.6$, where the impact of exotic DM models on matter distribution is less affected by non-linearities, with greater statistical power than previously possible.

The spectrographs used by DESI have a resolution between R=2000 and R=5100, up to a factor of 2 better than SDSS, and the survey is the first to apply the spectro-perfectionism approach \cite{Bolton_2010} to optimize the data reduction accuracy. The increased spectral resolution enables structures of smaller physical size to be probed with unprecedented precision: previous samples of similar or better resolution have been constrained to $\lesssim 1000$ objects in total \cite{Murphy2019,Omeara_2021,Karacayli_2022}.
DESI uses a Random Forest machine learning algorithm for the quasar target selection, and a combination of a spectral fitting classifier (Redrock\citep{Redrock}) and a deep convolutional neural network classifier (QuasarNET\citep{busca2018quasarnet}) to distinguish between quasars, galaxies and stars, measure the objects' redshifts and generate a quasar catalog. 
Using this catalog, a software package (picca\footnote{\url{https://github.com/igmhub/picca}}) is used to statistically determine the quasar continuum, create a Ly-$\alpha$ absorption catalog, which contains the transmission fields as the fraction of observed flux over the mean absorbed continuum, and calculate the correlation functions and power spectra.

\section{Observational Signatures of Dark Matter Physics \label{sec:predictions}}

The standard cold dark matter with dark energy model  ($\Lambda$~CDM) has been extremely successful at matching theoretical predictions from numerical simulations with the observed distributions of galaxies on large scales. However inconsistencies persist when the predictions from simulations are compared to observational data on small scale. This is the so-called ``small-scale crisis'' of CDM. This and some of the other key predictions of $\Lambda$CDM that can be probed by DESI and could yield insights into the nature of dark matter are described below.

\paragraph{The ``cusp-core'' problem'' and the ``diversity of rotation curves'':} 

Simulations carried out with CDM predict universal dark matter density profiles with steeply rising central density cusps \citep{NFW97}. However models of the central stellar and gas kinematics in dwarf spheroidal galaxies, low surface brightness disks and gas-rich dwarf galaxies overwhelmingly favor cored density profiles, although exceptions exist. This is referred to as the ``cusp/core problem'' \citep[for a review see,][]{Bullock_Boylan-Kolchin_2017}.

The CDM density profiles are almost identical to the profiles predicted by warm dark matter  \citep[WDM, e.g. a sterile neutrino,][]{vogelsberger_2016_ethos}, except for halo masses below $\sim 10^{10}$~\msun\, \citep{Bose_coco}, but quite different from the cored density profiles that result from simulations with self-interacting dark matter (SIDM) \citep{Rochas_Peter_2013_SIDMcores,Tulin_Yu_2018} and fuzzy dark matter (FDM)~\citep{Schive2014,axgadget,Nori2021}.   The central density cores  observed in low surface brightness and dwarf galaxies are more consistent with the central density profiles of halos produced in SIDM simulations \citep{Rochas_Peter_2013_SIDMcores} than with the cuspy halos predicted by CDM and WDM simulations. Also SIDM more naturally produces the observed diversity of rotation curves  \citep{Oman_Navarro_2015}.

The ``cusp/core'' problem is significantly alleviated by considering feedback from bursty star formation \citep{pontzen_governato_12} which is able to produce cores in subhalos of mass $\lesssim 10^{10}$\msun~  \citep[for an alternative view see][]{bose_2019}.  However, classical and ultra faint dSphs ($M_*/M_{\mathrm{halo}} \leq 10^{-3}$) have formed so few stars over their lifetimes, that it is challenging to invoke baryonic feedback (e.g. stellar winds and supernovae) as the main mechanism responsible for drastically transforming their central dark matter distributions \citep{Lazar_etal_2020}. Cosmological hydrodynamical zoom-in simulations with different types of dark matter (CDM, WDM, SIDM and mixed DM, e.g. WDM with self-interaction, \citep{Fitts_etal_2019}) show that the addition of baryons substantially decreases differences between the simulations with different types of dark matter. Small differences  do exist however, e.g. baryons decrease the sizes of cores in SIDM and WDM+SIDM subhalos compared to SIDM-only simulations, but SIDM halos have lower central densities than CDM halos with baryons.  It has also been found \citep{Fitts_etal_2019}  that baryons in the SIDM and WDM+SIDM simulations result in a relation between circular velocity at the stellar half-mass radius $V_{1/2}$ and half mass radius $r_{1/2}$  that is broadly consistent with observations.  One of the greatest successes of the SIDM+baryons model is probably its ability to account for both the diversity of rotation curves \citep{Kamada_Kaplinghat_2017}, where the DM profile closely follows that of the baryons \citep{kaplinghat_2019,kaplinghat_etal_2020} and produce {\it ab initio}   the radial acceleration relation \citep{mcgaugh_etal_2016} - the relation between total acceleration and the acceleration due to baryons. {\it DESI's large field of view, high sensitivity and excellent radial velocity precision will enable us to obtain radial velocities for significant samples of stars in dwarf spheroidal and ultra faint satellites of the MW enabling us to set better constraints on their density profiles} (see \S~\ref{sec:dwarfs} for details.)

\paragraph{The total mass and number of satellites of the Milky Way:} 

Substantially fewer satellites are found around MW-like galaxies than predicted by dark matter-only $\Lambda$CDM simulations \citep[the ``missing satellites problem'', e.g.][]{moore_etal_99, klypin_etal_99}. On the other hand, WDM, and a variety of other alternatives to CDM, result in a suppression of small-scale power at early times and consequently  significantly supresses the number of subhalos of mass $\leq 10^8$\msun \citep[e.g.][]{Bose_coco,Newton_2021_WDM}, and similarly for the case of FDM~\citep{Schive2016,Schive2018}. The  subhalo mass function in SIDM simulations is largely indistinguishable from CDM simulations and the only models that lead to a difference relative to CDM have velocity-independent cross-section of $\sigma/m \gtrsim 10\,$cm$^2$g$^{-1}$, which has already been ruled out by  observations of clusters of galaxies, like the Bullet Cluster \citep{Randall_2008}. However, baryonic feedback from energetic stellar winds and supernovae can also decrease the number of observed dwarf satellites by blowing out the baryons and hence quenching subsequent star formation in smaller subhalos. This could result in a population of (nearly or completely) dark subhalos whose presence could be inferred by their gravitational impact on visible structures in the Galaxy (such as stellar streams) or gamma ray emission if composed of WIMPs. In external galaxies strong lensing can also be used to infer the presence of dark subhalos \citep{Vegetti_2010}. $\Lambda$CDM simulations also predict that the most massive satellites of Milky Way mass galaxies are more massive and denser on average than the observed satellites  \citep[referred to as the ``too-big-to-fail problem'',][]{Boylan-Kolchin_2011_TBTF}.

Uncertainty in the total (virial) mass of the Milky Way greatly affects our ability to assess whether or not the observed number of satellites and their masses are consistent with predictions from $\Lambda$CDM. Although the mass of the Milky Way is a fundamental quantity of interest for comparisons with predictions from cosmological simulations, it is surprisingly poorly constrained. The availability of distances, radial velocities and even proper motions for huge numbers of individual stars, globular clusters and satellite galaxies has led to numerous efforts to determine the Milky Way halo parameters. Recent reviews give estimates of the mass of the Milky Way  in the range 0.55-2.62$\times 10^{12}\msun$ \citep[e.g.][]{Bland-Hawthorn_Gerhard_2016,wang_20}. Methods using tidal streams, give a similarly large range of masses. The most recent measurements (based on {\Gaia\,} and/or {\it HST} data) demonstrate that despite the availability of high quality data, the mass of the Milky Way is still uncertain by a factor of about two.
These uncertainties affect our ability to determine how severe the ``missing satellites problem'' and ``too-big-to-fail problem'' really are. {\it DESI's radial velocity data will enable us to decrease the uncertainty in the mass from its current value of 200\% to $\lesssim 25$\% using tracers out to at least 100~kpc, which will result in improved constraints on a range of DM models}  (see \S~\ref{sec:global} for details).

\paragraph{The shapes of dark matter halos:} 

Dark matter halos in CDM-only simulations have overwhelmingly triaxial shapes \citep[e.g.][]{dubinski_carlberg_91,jing_suto_02, Rossi_2011,Prada2019} with flattening (short/long axis ratios $c/a \sim 0.6$ and intermediate/long axis ratios $b/a \sim 0.8$). Triaxiality correlates with halo mass, with triaxiality decreasing with decreasing halo mass \citep{Allgood_etal_2006}. Dark matter-only simulations with SIDM also predict triaxial dark matter halos, but the degree of triaxiality depends on the self-interaction cross section ($\sigma/m$) 
\citep{peter_rocha_2013_SIDM_shapes}. WDM halos (made e.g. of sterile neutrinos) resemble CDM halos on Milky Way mass scales but are rounder on smaller scales \citep{Bose_coco}. Baryons tend to make the distinctions between different types of dark matter less clear. In CDM-only simulations, baryons make halos more oblate-axisymmetric within the inner third of the virial radius, becoming more triaxial at large radii \citep{Kazantzidis2004,Zemp2012,Chua2019}. Recent SIDM simulations of Milky Way mass halos with different interaction cross-sections show that the addition of baryons results in halo shapes that are similar to CDM halos \citep{vargya_etal_2021}. However the {\it radius} at which the shape of the stellar halo deviates from the shape of the total matter distribution is a sensitive indicator of the SIDM self-interaction cross-section $\sigma/m$. {\it By obtaining radial velocities for Blue Horizontal Branch (BHB) stars, RR Lyrae, and luminous red giants out to $\sim 100$~kpc, DESI will be able to measure the radial change in shape of both the DM halo and stellar halo, providing new constraints on the nature of the DM particle} (see \S~\ref{sec:global} for details.) 

\paragraph{Angular momentum content of dark matter halos:}
Halos acquire a net angular momentum through tidal torques during growth in the linear regime \citep{Peebles_1969, Doroshkevich_1970, White_1984}.  This angular momentum can subsequently be modified and redistributed by mergers. It is now common to measure of halo angular momentum by defining a dimensionless spin parameter  $\lambda = J/(\sqrt{2}M r_{\mathrm vir}v_{\mathrm vir})$, where $J$ is the total angular momentum of a galaxy,  $M$ is the total mass, and $r_{\mathrm vir}$ and $v_{\mathrm vir}$ are the virial radius and velocity, respectively \citep{Bullock_etal_2001}.  
In cosmological simulations it has been found that the angular momentum of the dark matter halo is correlated with the angular momentum of  stellar disk as well as  the angular momentum of the stellar halo  \citep{obreja_etal_2022}, enabling dark matter spin to be measured via stellar kinematics. 
The angular momentum of merger remnants can be either in the form of streaming motions of individual particles  or figure rotation (or tumbling). It has been long known from CDM cosmological $N$-body simulations that $\sim$90\% of dark matter halos are  triaxial and can tumble  with very slow pattern speeds \citep{dubinski_94_shapes,bailin_steinmetz_04,bryan_cress_07}. 
Unlike several other properties of DM halos, both $\lambda$ and pattern speed are independent of mass. The measurement of these quantities in the MW will allow us to distinguish between particle models of dark matter and non-gravitational and non-Newtonian theories \citep{bailin_steinmetz_04}.
{\it DESI's measurement of radial velocities and precision spectrophotometric distances for halo stars, in  conjunction with proper motions from \Gaia\, (and Roman's HLS) will enable some of the first measurements of the angular momentum content (halo spin and pattern speed) of the MW halo} (see \S~\ref{sec:global} for details.)


\paragraph{The small-scale Ly-$\alpha$ forest power spectrum}

The primordial power spectrum is ideally probed by the Ly-$\alpha$ forest on scales ranging roughly an order-of-magnitude around $k\sim 1\,h^{-1}$~Mpc. The density fluctuations as probed by the Ly-$\alpha$ forest are still in a mildly non-linear regime since the forest signal is generated by the IGM in relatively empty regions of the cosmic web, and it is measured at redshifts $z>2$. 
This makes the Ly-$\alpha$ forest a unique probe of primordial small-scale density fluctuations where signatures of alternative DM models arise. A small-scale cut-off in the early matter power spectrum is predicted in several DM scenarios beyond CDM. In particular, it is a consequence of the velocity dispersion of non-thermal WDM particles, e.g. sterile neutrinos produced by resonant transitions \citep{Sterile1999,Sterile2001,Sterile2013,Sterile2015,Sterile2016,SterileEnqvist:1990ek} or parent particle decays
\citep{Sterile2006, Sterile2008,
Sterile2014, Sterile20152, Sterile2016Decay, sterile2016Decay2}; 
or of the quantum pressure produced by FDM particles, e.g. axion-like particles, with extremely small masses $m\sim 10^{-22}\text{eV}$
\citep{Fuzzy2000,Fuzzy2013,axion2013,Hui_2017} (for an early discussion of the cosmological consequences of ultra-light boson particles for dark matter see also~\cite{Sahni2000,Goodman2000,Matos2001,Arbey2002}) ; or of DM interactions described by e.g. the  Lagrangian-based Effective theory of structure formation (ETHOS) \citep{ETHOS2016,ETHOS20162}; or oscillations induced by interactions in the dark sector \citep{Murgia2017,Hui_2017,Bose_2019b,Schaeffer2021}. The various beyond-CDM scenarios predict different shapes for the power spectrum
suppressions: for example FDM and ETHOS predict oscillations at very small scales (for FDM see for instance~\cite{Medellin2021,Lague2021}), contrarily to WDM. However, the Ly-$\alpha$ forest is not sensitive in general to these detailled features, and a general $3$-parameter formula can be used to approximate the transfer function of those DM scenarios \citep{Murgia2017}, with a precision adapted to Ly-$\alpha$ observations. Besides, using high-resolution quasar spectra, different studies have detected a cut-off in the small-scale power spectrum~\citep{viel_how_2007,viel_warm_2013,irsic_first_2017} coherent with the signature of alternative DM models such as WDM or FDM, preventing the growth of small-scale structures. However, a similar cut-off in the flux power spectrum is produced by well established astrophysical effects,  such as the thermal Doppler broadening due to the instantaneous temperature of the IGM, and the smoothing of the gas distribution due to the photo-heating of the gas by the first sources of reionization and the ongoing emission from galaxies and quasars \citep{Garzilli:2015iwa,Garzilli:2018jqh,Walther2019}. To break this degeneracy one can exploit the different redshift evolution of those effects, which requires particularly accurate modeling of IGM physics in hydrodynamical cosmological simulations, the only way to provide predictions that match the scales and precision achieved by the data (see \S~\ref{sec:sim_lya}).
\textit{Due to the strong improvement in resolution and the observation of fainter quasars (see \S~\ref{sec:Lya-survey}), DESI will be able to measure the Ly-$\alpha$ power spectrum at scales as small as 5 Mpc.h$^{-1}$ at redshifts as high as $z$=5 enabling significantly tightened constraints on alternative DM models.}

\section{Dark Matter Constraints from DESI 
\label{sec:DM_desi_constraints}}

\subsection{Constraints from the Milky Way Survey \label{sec:MWS}}

Our unique vantage point within the Milky Way enables us to directly test several different predictions of cosmological simulations described in Section~\ref{sec:predictions}. DESI in concert with {\Gaia\,} and future proper motion surveys can create the large, homogeneous datasets required for testing various dark matter properties. These data will enable us to set competitive constraints on both the global and small-scale properties of the dark matter distribution within the Milky Way and its satellites, thereby allowing us to place the Milky Way's measured properties and its evolutionary history in a cosmological context. 
Suites of high spatial and mass resolution (zoom-in) cosmological simulations with baryons run with different dark matter scenarios, will be needed to assess the likelihood of obtaining the observed global properties of the Milky Way halo and the ensemble properties of its subhalo population with each type of dark matter particle. 
The global and small-scale constraints on dark matter that DESI will contribute to are described below.

\subsubsection{Global halo constraints \label{sec:global}}

DESI will yield an unprecedented sample of radial velocities for halo giants (hundreds of thousands to millions, depending on the density confirmed by DESI spectroscopy), as well as stellar standard candles such as RR Lyrae and BHB stars, out to $\sim 100$~kpc. These radial velocities with {\Gaia\,} DR5 proper motions, and spectrophotometric distances  from DESI (with $\sim 10$\% precision) will yield the 6D phase-space coordinates for field halo stars that are needed to determine the cumulative mass of the Galaxy out to 100~kpc with an accuracy of a few percent \citep{Sanderson_2015,Bonaca2018ColdStreams} to 25\% \citep{rehemtulla_etal_2022}, significantly reducing the current uncertainty of 200\%. In addition, accurate measurement of the velocities of halo stars in the wake produced by the LMC \citep{garavito-camago_19,conroy_etal_2021} will allow us to determine the masses of the LMC and the Milky Way and to correct for the effects of disequilibrium caused by the LMC's tidal field \citep{Deason_2021,correa-magnus_vasiliev_2021}.

These 6D phase-space coordinates for field halo stars will also enable the measurement of the 3D shape of the dark matter halo and its variation with radius, in order to enable comparisons with cosmological simulations with different types of dark matter. Sophisticated stellar distribution function modeling methods have recently been developed and validated with mock data for this purpose \citep{hattori_etal_2021}. The tests with mock  Gaia DR2 data generated from Milky Way like galaxies in cosmological hydrodynamical simulations show that the flattening (short axis to long ratio $c/a$) can be measured to a precision of $\Delta(c/a) \pm 0.15$ even without radial velocities. Adding radial velocities decreases the uncertainty to $\Delta(c/a) \pm 0.1$. The current measurement of $c/a = 0.96$ obtained with \Gaia\, RR Lyrae stars  \citep{Wegg2019,hattori_etal_2021} are already mildly in tension with $\Lambda$CDM simulations which predict $c/a = 0.8\pm 0.1$ \citep{Chua2019}. DESI will make it possible, for the first time, to measure the variation in shape of both the dark matter and stellar halos, enabling comparisons between the MW and cosmological simulations which predict differences between SIDM and CDM \citep{vargya_etal_2021}.

Stellar tidal streams have also been used to derive the shape of the Milky Way's dark matter halo. Since tidal streams deviate only slightly from the orbits of their progenitors, they provide an important way to constrain the orbits of objects in the global potential of the Galaxy.  Previous models of the Sagittarius stream \citep{helmi_2004, fellhauer_etal_2006, law_majewski_10} have yielded highly discrepant results, in large part due to the absence of precision kinematic data (proper motions and line of sight velocities) along the entire stream. Although {\Gaia\,} provides proper motions along the entire stream, the proper motion uncertainty is $\gtrsim 10-20$~\kms  at $50-100$~kpc where much of the stream lies. The best recent model of the Sagittarius stream \citep{vasiliev_tango} used {\Gaia\,} DR2 proper motions and limited radial velocity data  to quantify the effects of the LMC on the MW and the Sagittarius stream but did not attempt to constrain the shape of the halo. Since the Sagittarius stream occupies a significant portion of the sky, the large area and depth of DESI will, over the course of the 5-year survey, significantly increase the sample of stars with radial velocities with precision of $\lesssim 10\kms$ out to 100~kpc  along  very large contiguous lengths of the Sagittarius stream, enabling us to set stronger constraints on the global variation in halo shape. 
Constraining the orbits of multiple stellar streams simultaneously can serve to set stronger constraints on the global mass distribution (the shape and density profile of the halo) \citep{bovy_16}. It has been shown that the availability of full 6D phase-space coordinates for about a dozen stream can yield constraints of a few percent on the total halo mass (within the radius occupied by the streams) \citep{Sanderson_2015,Bonaca2018ColdStreams}. {\it DESI is the only current US survey with the field of view large enough be able to efficiently map the radial velocities of stars along the long streams (tens of degrees across the sky) that are most useful for global halo modeling.}

Finally,  although it has been long known from CDM cosmological simulations that $\sim$90\% of dark matter halos exhibit steady figure rotation (tumbling) \citep{dubinski_94_shapes,bailin_steinmetz_04,bryan_cress_07}, 
it has only recently been shown that it is possible to measure (or at least set an upper limit) on the pattern speed of figure rotation of the Milky Way halo using the phase-space distributions of stars in long halo tidal streams like the Sagittarius stream \citep{valluri_etal_2021}. The smallness of the pattern speed predicted ($0.15h -0.8h\kmskpc \sim 9-45^\circ \mathrm{Gyr}^{-1}$) implies the need for  precise 6D phase-space data over a significant portion of the stream, implying that such a measurement is not possible in any external galaxy. While there are currently no published studies of figure rotation of dark matter halos arising from different types of dark matter particles,  even the detection of figure rotation will allow us to rule out  \citep{bailin_steinmetz_04} non-standard gravitational theories such as MOND \citep{Milgrom_1983,Sanders_1990,Milgrom_2002}. The first measurement of the spin  parameter $\lambda$ of the Milky Way halo has also been made recently from {\Gaia\,} measurements of the kinematics of Milky Way halo stars \citep{obreja_etal_2022}. This measurement is based on correlations between the angular momentum content of the stellar halos and dark matter halo spin in $\Lambda$CDM   hydrodynamical simulations. DESI will greatly increase the sample of stars and the fraction of the volume of the halo from which the spin of the stellar halo can be measured, enabling us to improve the measurement of halo angular momentum.

\subsubsection{Constraints  on dark matter subhalos from stellar tidal streams \label{sec:streams}}


In addition to being powerful probes of the global dark matter distribution in the Milky Way, stellar streams originating from globular clusters (GC) are powerful probes for the dark matter distribution on smaller scales. GCs have small velocity dispersions ($\lesssim 3-5\kms$), consequently tidal streams from GC progenitors are dynamically ``cold" and narrow. This makes them sensitive probes of  perturbations due to gravitational interactions with small-scale perturbers such as dark subhalos \citep{Erkal_Belokurov_2015_gap1, Erkal_Belokurov_2015_gap2,Erkal2016_gaps}, since such interactions produce gaps, density irregularities and off-stream structures that are easier to observe in the otherwise thin streams.  In particular, since subhalos of mass less than $10^8$\msun\, are predicted to exist in Milky Way like halos in the standard $\Lambda$CDM cosmological scenario, but are hard to detect due to their lack of baryons, observations of such gaps provide a crucial test of the clustering of dark matter at mass scales below the threshold of galaxy formation. Gaps have already been observed in streams like the Palomar 5 stream \citep{odenkirchen_2001} and the GD1 stream \citep{Grillmair_Dionatos_2006}. Models for the formation of the gap in the GD1 stream for instance require a compact object of $\sim 10^7$\msun~\, with much higher density than expected for $\Lambda$CDM subhalos \citep{Bonaca_spur_2018}. The measurement of the velocities of  stars in the vicinity of gaps can lead to stronger constraints on the orbit, mass, and velocity of a possible perturber \citep{Erkal2016_gaps}. More precise predictions on the locations of the perturbers based on DESI data could also lead to the discovery of brand new targets for gamma-ray dark matter searches \citep{2021JCAP...11..033M}. It has been shown that analyzing the power spectrum of density fluctuations in a stream can enable constraints on the mass function of dark subhalos, since subhalos of different masses affect the density structure of streams on different scales \citep{Bovy_etal_2017}. Recent observations of gaps in the GD1 stream have been used to infer the total abundance of dark subhalos normalized to standard CDM \citep{Banik_2021b}, and can possibly enable constraints to be set on the nature of the dark matter particle \citep{Banik_2021}.

At least half of the Milky Way's GCs are thought to have formed in external dwarf galaxies as massive star clusters that were subsequently accreted into the Galaxy. Thin and dynamically cold GC streams  have recently  been shown to be sensitive probes of the  central dark matter distributions in the dwarf galaxies that accreted GCs were born in. 
GCs accreted within cosmologically motivated subhalos (hosting dwarf galaxies) experience a phase of ``pre-accretion tidal evolution''  within their parent subhalo \citep{Carlberg_2018,Malhan_etal_2021,Qian_etal_2022}. After this pre-accretion phase, the GC and its host dwarf galaxy are accreted by the Milky Way and the dwarf galaxy is tidally disrupted, while the GC stream continues to evolve within the Milky Way potential. 
This two-phase evolution produces tidal streams that are more highly structured than GC  tidal streams that evolve in the MW alone. Accreted GC streams comprise a thin, linear component plus a broader ``cocoon'' component as  well as off-stream structures (``spurs'') and can have numerous gaps and branches.  Cocoons, spurs and gaps have now been detected with {\Gaia\,} photometry and proper-motions in several GC streams \citep{Malhan_etal_19cocoon, Bonaca2019Jhelum}. It has been  shown recently
that quantities that measure the ``hottness'' of accreted  GC tidal streams (physical width, line-of-sight velocity dispersion, tangential velocity dispersion, 
and angular momentum spread) are sensitive to whether the parent satellite from which the GC was accreted had a cuspy or cored dark matter halo \citep{Malhan_etal_2021, Malhan_etal_2022}. These simulations have found that stellar streams formed in cored DM halos are systematically narrower (in physical width) and have lower velocity dispersion (both line-of-sight and tangential) than those formed in cuspy halos \citep{Malhan_etal_2021, Malhan_etal_2022}.   A comparison of  simulations with existing {\Gaia\,} proper motions in about 7 streams already imply that many observed GC streams are too cold to be consistent with formation in cuspy CDM halos of  mass $\gtrsim 10^9$\msun. DESI will enable the measurement of line-of-sight velocity dispersions in many more streams, including those too distant for making a measurement with Gaia proper motions. 

This novel probe of the central density profiles of dwarf galaxies is a powerful complement to density profile measurements made with stellar kinematics in dwarf galaxies (see Section~\ref{sec:dwarfs}). {\Gaia\,} has discovered dozens of Milky Way GC streams and measured the proper motions of their stars \citep{Malhan_etal_2018_tributaries,ibata_etal_2019}. Many of these lie within the DESI footprint and DESI's large field-of-view and high survey speed  will enable the radial velocity and stellar abundance measurements needed to obtain  constraints on the density profiles of the parent halos of accreted GCs.  
Although impacts on GC streams from dark subhalos and pre-accretion tidal evolution of accreted GC streams can both produce gaps and off-stream features like those observed in the GD1 stream, kinematics of stars along substantial sections of tidal streams can distinguish between these two scenarios. 

Radial velocities are a crucual complement to proper motion surveys. DESI,  with its large field-of-view and survey area, is currently the most efficient way to obtain radial velocities and determine stream membership for a significant number of stars in stellar streams.  {\it In other words, a spectroscopic survey like DESI is crucial for leveraging the power of GC streams to probe the nature of dark matter.} 



\subsubsection{Constraints on disk dark matter from galactoseismology \label{sec:disk}}
Direct dark matter detection experiments rely on precise knowledge of  the dark matter density in the solar neighborhood. Traditionally this has been estimated by using the stellar number density and velocity distributions (typically perpendicular to the disk) or modeling the rotation curve  \citep[for recent reviews see,][]{Salucci_2019,desalas_widmark_2021}.  Some cosmologically motivated simulations have predicted that a dark disk, co-planar with the stellar disk, could exist in the Milky Way, formed either from the accretion of satellites \citep{Read2008, Purcell2009} or from a subdominant species of dark matter particle with much stronger interactions than the collisionless WIMP. Such a particle could interact dissipatively forming a rotationally supported disk \citep{Fan_2013a, Fan_2013b}. While searches for such a ``dark disk'' have been largely unsuccessful \citep{Ruchti_2015}, this lack of success could in part be due to the assumption that the disk is in steady state \citep{Widmark_2021a}.

Spectroscopic surveys of the stellar disk from SEGUE, RAVE, LAMOST revealed that the local stellar disk of the Milky Way exhibits bending and breathing modes perpendicular to the midplane within a few kpc from the sun \citep[for a review see][]{Widrow2018}. Recently 6D phase-space data from the {\Gaia\,} satellite has led to the discovery of even more complex kinematic substructures in the stellar disk in form of  ``phase-space spirals'' and ridges \citep{Antoja_2018}. The presence of the phase-space spiral in stars of all stellar ages \citep{Laporte_2019,Bland-Hawthorn_2019}, and the detection of the spiral at different radii  and azimuthal angles in the disk \citep{Xu_2020} suggests that these features are a large scale (disk-wide) response that is possibly the consequence of the passage of the Sagittarius dwarf through the stellar disk. 

It has recently been shown that the phase-space spiral, an inherently non-equilibrium feature, especially its morphological structure (its width, flattening and degree of winding), can be used to set extremely tight constraints on the potential of the Galaxy, especially the potential associated with a thin dark disk \citep{Widmark_2021a,Widmark_2021b}. These authors used {\Gaia\,} EDR3 kinematics to constrain the surface density of a thin dark disk with a scale height of less then 50~pc to be 5$\msun$pc$^{-2}$ much tighter than the constraints placed by previous steady-state models that also used {\Gaia\,} data \citep{Schutz_2018}, essentially ruling out the existence of a ``dark disk'' and therefore more dissipative  species of dark matter particles.

Although {\Gaia\,} has been crucial for the discovery of the phase-space spiral and the models arising from its discovery, {\Gaia's} Radial Velocity Spectrograph will only obtain radial velocities for stars brighter than $G=16$. {\Gaia's} end of mission radial velocity precision for stars at $G=16$ is expected be 10-20~km~s$^{-1}$ (depending on stellar type).  Furthermore the {\Gaia\,} database will not include stellar abundances. While the DESI MWS main survey is primarily targeting stars at $|b|>30^\circ$, the Backup Program (conducted during poor seeing and very bright conditions) will  get much closer to the disk plane targeting stars with $16<r<18$ and $|b|>7^\circ$. The current performance of DESI during the Backup Program suggests that we will be able to obtain $\sim 1-10$km~s$^{-1}$ precision for $r=16 - 18$ respectively. This will enable constraints to be set on the dark matter density in the disk out to much larger distances, thereby setting stronger constraints on exotic dark matter particle scenarios.

\subsubsection{Constraints on dark matter distributions of Local Group dwarf galaxies \label{sec:dwarfs}}

Dwarf spheroidal galaxies and ultra faint dwarf galaxies are amongst the most dark matter dominated, most metal-poor and oldest stellar systems in the Universe representing the low end  of the galaxy luminosity function \citep{Simon_2019ARAA}. The fact that they contain very few stars may imply that these systems are some of the only places in the Universe where one can  directly probe the distribution of dark matter on small scales. While a few dwarf spheroidal galaxies have been found to possess cuspy DM profiles e.g., Draco \citep{Jardel2013DracoCusps, Read_2018_draco} and Sculptor  \citep{amorisco_evans_2012}, the majority of galaxies modeled to date strongly favor DM distributions with shallower, almost cored, inner density profiles.  Cored central density distributions have now been measured in low surface brightness disks \citep{Moore1994, Burkert1995, deBlok2001_cored}, and various dwarf satellites of the Milky Way, e.g. Ursa Minor \citep{Kleyna2003UrsaMinorCore}, Fornax \citep{Goerdt2006Fornax, Walker_Penarrubia_2011, Cole_2012_fornax, Pascale2018}, NGC 6822 \citep{Weldrake2003} and Eridanus II \citep{Contenta2018EridanusII}. 
However, despite the strong claims of cores in the literature, the current measurements of central density profiles are generally based on very small numbers (hundreds to a few thousand) of line-of-sight velocities in each galaxy and strong assumptions about the internal velocity anisotropy. The velocity dispersion in these galaxies is of order 2-5~\kms and the precision of the line-of-sight velocity measurements is of the same order resulting in fairly large uncertainties in the central mass profiles and velocity anisotropy distributions \citep{kowalczyka,Read2021GaiaChallenge}. Furthermore, although dSphs are generally modeled as spherical, they are observed to be elongated (prolate or triaxial) and this results in significant uncertainties in both the density and anisotropy profiles \citep{kowalczykb,Genina18}. While proper motions can help to reduce the uncertainties \citep{Strigari_etal_2007, Read2021GaiaChallenge} the precision of proper-motions from the {\Gaia\,} satellite are currently inadequate to distinguish between a cusp and a core in dwarf galaxies like Sculptor \citep{strigari_etal_2018}. For many dSph galaxies, the 2$\kms$ tangential velocity precision needed to obtain tight constraints on the central density profiles will become available following several years  of observations with the Roman Space Telescope. 

In addition to the primary DESI-MWS, DESI will also conduct a secondary program to observe all the dwarf spheroidal galaxies and ultra-faint dwarf galaxies within DESI footprint. Most dwarf spheroidal and ultra faint dwarf satellites of the Milky Way comfortably fit within DESI's 3$^\circ$ field of view. The large numbers of fibers is ideally suited to obtaining radial velocities for a significant fraction of stars in these systems. The candidate member stars are selected using the proper motion and parallax information from Gaia EDR3, together with the photometry from the DESI Legacy Survey DR9 \citep{Dey_2019_DESILS}, and will be observed during the bright time survey for stars at $16<r<20$ and during the dark time survey for stars at $19<r<21$. We expect that with multiple passes surveying the sky in the 5 year DESI survey, we will observe most of the member stars in the ultra-faint dwarf galaxies, and a significant number of stars (a few thousand stars per galaxy) in the dwarf spheroidals. 

Radial velocities of stars in the cores of dwarf galaxies in conjunction with $\sim 25~\mu$arcsec proper motions (expected from the Roman Space Telescope) will provide precision measurements of their central density profiles, addressing the cusp/core problem and enabling indirect dark matter searches to place more robust and stringent constraints, not only on CDM WIMPs using gamma-ray data \citep{Ackermann:2015zua,2015ApJ...809L...4D,Fermi-LAT:2016uux,2016PhR...636....1C,2017MNRAS.466..669C} but also on WDM particle candidates -- such as sterile neutrinos or gravitinos -- from observations in the X-ray and gamma-ray energy bands \cite{2014PhRvD..90j3506M,2014PhRvD..89b5017H,2021PhRvD.104b3022A}, axion dark matter with radio data \cite{2018PhRvD..98h3024C}, interacting and self-interacting dark matter models \citep{2014MNRAS.445L..31B,2018PhRvD..98d3017B} and possibly other, even more exotic scenarios \citep[e.g.][]{2021PhRvD.103l3028W}. 

Despite substantial efforts to obtain spectra of stars in the centers of these systems, very few spectroscopic observations have been taken in their outskirts, mainly due to the limited field of view of existing spectroscopic facilities. Recent work suggests at least some of these galaxies may be surrounded by significant low surface brightness structures. Such structures have been predicted by simulations for certain types of dark matter even for galaxies that are not presently losing significant mass through tidal stripping \citep[e.g.][]{WangMeiYu2017}. The primary DESI MWS  will provide a wide but sparse sampling of several members in the outskirts of these galaxies.  Dedicated follow up in secondary programs
will enable the study of large-scale metallicity gradients and tidal effects. With both proper motions and radial velocities dynamical models may be able to set constraints on the 3D shapes of dwarf satellite galaxies \citep{An_Koposov_2022} enabling comparisons with simulations.


\subsection{Dark Matter Constraints from the Ly-$\alpha$ Forest}

Until now, the major statistical tool to constrain DM scenarios with Ly-$\alpha$ has been the one-dimensional Ly-$\alpha$ transmission power spectrum, $P_{\rm 1d}(k)$, introduced in \citep{Croft1998} with first applications to constrain DM properties in \cite{Seljak:2006qw}. This statistical property of the Ly-$\alpha$ forest is obtained by computing the power spectrum of Ly-$\alpha$ forest transmission in individual quasar lines of sight and then averaging over a quasar sample, ignoring 3D correlations. This allows the structure of matter fluctuations on small scales to be probed without being limited by the sampling of objects on the sky. The smallest scales that can be accurately probed are instead determined by the resolution of the spectrograph, the S/N of the dataset as well as astrophysical contaminants e.g. in the form of metal absorption systems.

Using large scale structure surveys, $P_{\rm 1d}(k)$ was measured for scales $0.001<k<0.02\,\rm s/km$ and redshifts $2.1<z<4.7$ from SDSS~\cite{mcdonald_lyman-alpha_2006}, BOSS~\cite{palanque-delabrouille_one-dimensional_2013}, and eBOSS~\cite{chabanier_one-dimensional_2019} with the latest measurement being based on $>45,000$ quasar spectra and reaching $\sim 1.5\%$ precision  for some redshifts. These measurements are complemented by smaller, high-resolution, high S/N~\cite{viel_warm_2013, irsic_lyman-alpha_2017,walther_new_2018,Karacayli_2022} $P_{\rm 1D}$ measurements from 8m-class telescopes, probing both smaller scales (higher k) and extending the redshift range to $1.8<z<5.4$, albeit at lower precision. Going to smaller scales and higher redshifts significantly improves constraints since alternative DM models have more impact at small scales.

The sensitivity of the high redshift Ly-$\alpha$ forest to DM is due to differences in the redshift evolution of typical physical scales shaping the signal. The DM models imprint a typical scale on the cosmic web at a fixed redshift (e.g. at decoupling) that is frozen in comoving distance units. At the same time, the physics of the intergalactic medium imprints two scales: the hydrodynamical response to reionization heating that increases with time elapsed since the heating occurred; and random motions due to finite temperature of the gas that is frozen in velocity units, resulting in increasing distance scale towards lower redshifts due to the Universe's expansion. Relative to the imprint of the baryonic physics, the scales dominating the dark matter sector become increasingly more important towards higher redshifts, see \citep{Rossi2017,Garzilli:2018jqh} for an in-depth discussion.

Additionally, due to the strong increase in quasar density, DESI for the first time will be able to measure small-scale Ly-$\alpha$ fluctuations both \textit{along and across} the line-of-sight (LOS) to  quasars, i.e. the 3D Ly-$\alpha$ power spectrum, $P_{\rm 3d}(k)$~\citep{AFR2019}. $P_{\rm 3d}(k)$ measurements will be a cosmic gold mine for Ly$\alpha$ DM studies for two reasons. First, it will add valuable constraining power at small scales, where we look for deviations from the $\Lambda$CDM model. Second, it will allow us to break degeneracies between the IGM temperature that smooths absorption along the LOS, and dark matter mass that smooths absorptions in all directions (but note that pressure broadening of the gas distribution due to past photo-heating is also a 3D effect and thus still degenerate with the WDM feature).

By combining high- and moderate-resolution surveys, the $P_{\rm 1d}(k)$ measurement was used to constrain WDM mass (thermal relic, sterile neutrino)~\cite{Viel2005,viel_how_2007,viel_warm_2013,baur_lyman-alpha_2016,yeche_constraints_2017,baur_constraints_2017,palanque-delabrouille_hints_2020,Garzilli:2021qos}. Most recent works agree on mass bounds of at least $m_X \geq 3.5 \rm keV$ when assuming a thermal relic type dark matter (with the strongest bounds $m_X \geq 5.3 \rm keV$ obtained when combining Ly-$\alpha$ samples~\citep{Irsic2017,palanque-delabrouille_hints_2020}) placing the Ly-$\alpha$ forest among the strongest probes of DM mass (see \cite{Enzi2021} for a recent comparison with other probes). We note that these constraints require careful modeling of IGM astrophysics in hydrodynamical simulations, which can relax constraints to $m_X \geq 1.9 \rm keV$~\citep{Garzilli:2021qos} when considering e.g. a particularly late reionization scenario.
 
Other models such as fuzzy dark matter predict a similar impact to WDM on the power spectrum of matter. The measurement of the $P_{\rm 1d}(k)$ also constrained these fuzzy dark matter models~\cite{irsic_first_2017,armengaud_constraining_2017,Linares2021}. In particular, using high and moderate resolution spectra, the authors in \cite{armengaud_constraining_2017} excluded the mass range of $10^{-22} < m_{a} < 2.9 \times 10^{-21}$ eV at 95 \% CL, although a possible self-interaction parameter of the axion particles remains unconstrained~\citep{Leong2019,Linares2021}.

DESI will significantly  improve upon eBOSS measurements. On the one hand, there will be a tremendous gain in statistical precision, going from tens of thousands of spectra to hundreds of thousands of spectra, and DESI will have one additional high redshift bin. On the other hand, DESI will double the $k$-range due to its improved spectral resolution, 2 times better than SDSS (see \S \ref{sec:Lya-survey}). 
As a result, DESI will be much less dependent on small high-resolution data sets
and will obtain the best Ly-$\alpha$ constraints on its own. As a benchmark scenario, we cons
ider a thermal relic WDM model, and compute $P_{\rm 1d}(k)$ predictions within this scenario using the hydrodynamical simulation grid of~\cite{2014JCAP...07..005B}. 
We define the sensitivity of a measurement as the 95\% CL bound on the WDM mass $m_X$, obtained on average when fitting mock CDM data to the WDM model.
Assuming $P_{\rm 1d}(k)$ is measured from SDSS data only, with uncertainties identical to those of eBOSS \cite{chabanier_one-dimensional_2019}, the sensitivity to the mass of WDM is $m_X > 2.3$~keV. 
On the other hand, we consider a DESI $P_{\rm 1d}(k)$ measurement with a factor of 4 increased statistics, and taking into account its improved resolution which permits measurements up to $k = 0.035$~s/km on average. In order to be conservative, we assume systematic uncertainties are similar to those estimated in eBOSS\cite{chabanier_one-dimensional_2019}, except that the improved resolution is taken into account.
We then find that the DESI-only sensitivity is $m_X > 3.6$~keV. This represents an improvement by a factor of 1.6 \textit{per se} compared to the eBOSS-only result. Moreover, the constraining power on IGM thermal parameters is improved by a factor of 2.6 on average: this will prove crucial to improving the robustness of constraints with respect to uncertainties in the IGM thermal modeling.


\section{Simulations for Dark Matter Science with DESI}

The vast majority of constraints currently available on the nature of dark matter come from comparisons between observations and cosmological simulations with $\Lambda$CDM, although there have been significant advances in the past decade on simulations with other types of DM (for a review see CF3-solicited white paper by Banerjee, Boddy et al.). Although there are clearly identifiable differences between the predictions from different types of dark matter, it has become increasingly clear over the past decade that it is much more difficult to distinguish between different DM scenarios when baryonic effects are considered \citep{Brooks_2019}. However, the next decade is expected to witness a substantial increase in the spatial and mass resolution of cosmological hydrodynamical simulations that will undoubtedly lead to significant improvements in the representation of DM halos both at high redshift and in the local Universe.

\subsection{Simulations for Milky Way and Local Group Science \label{sec:sim_MW}}

While modeling the data from DESI and other surveys will enable us to obtain constraints on the DM distribution in the MW and Local Group, these observationally derived parameters of the DM distribution need to be compared with simulations in order to set robust constraints on the nature of DM.  Understanding the full context of the Milky Way itself requires comparing to halos with similar accretion histories and satellite populations and understanding how such halos connect to the full distribution of predicted CDM systems. Example simulation suites that can enable such studies in $\Lambda$CDM are the ``Milky Way'' simulation suite (first introduced in \cite{Mao2015}) that consists of 45 resimulated host halos with $z = 0$ virial masses of $10^{12.1 \pm 0.03}\msun$, and a complementary set of 25 simulations (``Milky Way-est'' suite) drawn from the same parent box but chosen to have MW-like merger histories (e.g. a Gaia--Enceladus \citep{Belokurov2018,helmi_enceladus_18} analog merger at $z \sim 2$ then quiescent until the merger of an LMC analog with an infall time of less than 2 Gyr from $z = 0$; Buch et al. 2022 (in prep)). This combined simulation suite has already been used to successfully model the satellite populations observed by DES and Pan-STARRS \citep{DrlicaWagner2020, Nadler2020b, Nadler2021c, Mau2022}. In addition to the CDM simulations, three of the halos from the Milky Way mass suite have been resimulated using SIDM models, including velocity-dependent cross sections, from the same initial conditions \citep{Banerjee2020, Nadler2020a, Nadler2021a}. 
Comparisons between the SIDM and CDM simulations have already provided valuable insight on the relative subhalo population between the two models in the presence of the LMC \citep{Nadler2021a} and on the abundance, radial distribution, and density profile of SIDM subhalos \citep{Nadler2020a}. These works provide simulation-based predictions from different microphysical dark matter models for the internal subhalo dynamics that DESI will be able probe. In addition to resimulating the rest of the Milky Way-est suite using the SIDM prescription, future work will include using UniverseMachine \citep{Behroozi2019} to model star-formation histories for individual subhalos as well as comprehensive particle tracking down to the resolution limit of the simulations ($\sim 8 \times 10^7\msun$) to make detailed measurements of the profiles of both intact and disrupted subhalos in different dark matter models. There is also work being done to resimulate individual halos including realistic disk potentials. Combining these methods will allow for detailed predictions for dynamical clustering of disrupted substructure for different dark matter microphysics.

In addition to these on-going DM-only simulations that are ideal for investigating the subhalo populations, we need simulations that will improve up on existing cosmological hydrodynamical simulations which are achieving stellar mass resolution of $10^3-10^4\msun$ \citep[e.g., EAGLE, Auriga, APOSTLE, FIRE-2,][]{EAGLE_2015,Auriga2017, APOSTLE_2016,FIRE2}.

To meet the objectives laid out in this document we will need  a multi-pronged effort comprising the following:
\begin{itemize}
\item{A large scale cosmological hydrodynamical simulation effort designed to make dozens ($\sim 30-50$) of  Milky Way-like disk galaxies, possibly including a nearby satellite as massive as the LMC and a massive companion similar to  M31, that possibly had a significant accretion event  \citep[like Gaia-Enceladus,][]{Belokurov2018,helmi_enceladus_18} in the past $\sim 8-10$~Gyr. }
\item{This large suite of simulations should also be run with different types of DM and should start with identical (or as similar as possible) initial conditions.}
\item{The simulations should include realistic feedback prescriptions (e.g. from stellar winds, cosmic rays, and radiative mode feedback from AGN) designed to produce stellar halos that resemble the observed stellar halos of nearby MW mass disks. Currently most simulated stellar halos are too metal rich, too massive and have too many stars that formed in situ (i.e. were not accreted) \citep[e.g.][]{Cooper2010,Tissera2013,Pillepich_2015,monachesi_etal_2019,Merritt2020}.} 
\item{The simulations should produce stellar disks that resemble the global properties of Milky Way more closely. Currently stellar disks in cosmological simulations are too thick and too hot dynamically) \citep[e.g.][]{Grand2017,Grand_etal_aurigaia,Walo-Martin_2021_auriga_disks}; this is crucial for setting realistic constraints on the DM density in the disk.}
\item{Controlled cosmologically motivated simulations that evolve individual globular clusters (with realistic stellar populations) as well as populations of globular clusters in MW-like halos to generate thin tidal streams \citep[e.g.][]{EMOSAICS_2018,Li_Gnedin_2019,Kruijssen_etal_2019}. Such simulations should include impacts from  cosmologically motivated dark subhalos populations expected in different dark matter scenarios.}
\item{Tools to generate mock stellar populations with realistic stellar velocities properly accounting for observational selection functions of DESI and astrometric surveys (like Gaia, LSST, Roman-HLS) need to be applied in a consistent manner to all simulations and produce mock catalogs that are served on ADQL servers that can be queried in similar manner to observational catalogs, so that blinded studies can be easily performed.}
\item{Support for machine learning/deep learning methods that can be applied to DESI data (and other datasets) as well as mock catalogs to identify multi-dimensional correlations and patterns in the data. Given the numerous possible constraints that can be set on DM distribution in the MW, in the  event that no one observational probe is found to provide ``smoking-gun evidence'' of a specific DM particle, tools are needed to assess the likelihood that a given set of observational constraints is statistically consistent with predictions from simulations within a specific DM scenario.}
\item{Simulations of the formation and tidal evolution of dwarf galaxies (both classical dwarf spheroidal galaxies and ultra faint dwarf galaxies), and their tidal evolution within the MW halo and Local Group. }
\end{itemize}

The CF3-solicited white paper by Banerjee, Boddy et al. provides a comprehensive overview of the types of simulation effort needed for DM searches as a whole.
 
Regarding the creation of mock datasets from simulations there have been recent advances in this direction[e.g. Aurigaia  \cite{Grand_etal_aurigaia}, Ananke \cite{sanderson_18_ananke}, COSANG \cite{COSANG}]. These mock datasets yield realistic numbers of stars expected in different magnitude limited surveys as well as realistic representations of their phase-space distributions. However, the software to readily create mock datasets for different observational data releases (i.e. with different survey depths, data precision etc.) are publicly unavailable. In addition, not all these methods for creating mock data from simulations retain the kinematical properties of the constituent star particles and hence the ``stars'' they produce cannot be used as dynamical  tracers of the gravitational potential. Tests to ensure that the mock datasets are able to recover the true properties of the DM distribution, accounting for the effects of disequilibrium and cosmic variance are essential.

\subsection{Simulations for Ly-$\alpha$ science \label{sec:sim_lya}}

The thermodynamical state of the intergalactic gas giving rise to the  Ly-$\alpha$ forest signal is the result of the complex interplay between large-scale structure evolution driven by the dark-matter density fluctuations, and small-scale baryonic physics \cite{Hui1997}. Baryons in the intergalactic medium (IGM) trace dark matter fluctuations on Mpc scales, while on smaller ($\lesssim 100$ kpc scales) the $T\sim 10^4 K$ gas is pressure supported against gravitational collapse following the classic Jeans argument. Baryonic fluctuations are therefore suppressed relative to the pressureless dark matter and gas is pressure-smoothed or ``filtered'' on small scales. At any given redshift, the pressure smoothing scale depends not only on the prevailing pressure and temperature at that epoch, but also on the temperature history of the IGM \cite{McQuinn2016}. The density distribution of the baryons in the IGM is sensitive to the nature of dark matter in a similar, but importantly not identical way. The free-streaming horizon of dark matter particles lead to a 3D smoothing at high redshift (i.e.~in the initial conditions, in case of a simulation) that keeps decreasing as non-linearities increase. Therefore, the small-scale suppression in non-CDM dark matter models moves from very large scales to smaller scales at progressively lower redshifts. The major task of simulations in the context of constraining dark matter properties is thus to allow decoupling of reionization heating impact from the nature of dark matter on the  Ly-$\alpha$ forest statistics.

At first order, we thus require hydrodynamical simulations that accurately model the gravitational collapse of baryons and dark matter, and physical effects contributing to the gas pressure support, most notably radiative cooling and photoionization heating from the UV background. Studies have
shown that to produce converged  Ly-$\alpha$ 1-dimensional power spectra on the range of scales and redshifts produced by DESI, we need simulations of at least $\approx$120 Mpc with a $\approx$30 kpc resolution \cite{Lukic2015, rossi2020,Walther_2021}. As the the IGM  primarily probes densities around the mean density, i.e.~$0.1 \lesssim \Delta \lesssim 10$ depending on the redshift \cite{Lukic2015}, the above accuracy constraints translate into a minimal requirement of $4096^3$ cells for unigrid codes (e.g.~Nyx \cite{Almgren_2013, Sexton2021}, RAMSES \cite{Teyssier_2002}) or $8192^3$ particles for Lagrangian codes (e.g.~Gadget \cite{Gadget} and its variants). Larger numerical requirement on Lagrangian codes are stemming from the fact that at higher redshifts, $z>3$,  Ly-$\alpha$ forest originates mostly from underdense regions difficult to accurately sample with Lagrangian tracer particles (Chabanier et al.~2022, in prep.).

Recent improvements in high-performance computing facilities as well as cosmological hydrodynamical codes allow running simulations of this size. Note that to allow cosmological analysis, we do not only need a single simulation, but grids of simulations exploring the cosmological and astrophysical parameters allowing accurate predictions of the  Ly-$\alpha$ forest power spectrum. Those grids are used to train a fast interpolation scheme (emulator) for cosmological parameter inference \cite{Rogers2019, Takhtaganov2021}.  This is needed to bridge the gap between the need of millions of model evaluations we need to get accurate posterior probabilities on parameters of interest and the tens of simulations we can actually run. In effect, the emulator is used as a fast interpolation scheme between the actually simulated grid points in the final inference \cite{Heitmann2006}.  We foresee requiring tens of hydrodynamical simulations to achieve sufficient accuracy on this interpolation \cite{Pedersen_2021, Walther_2021}.

While the IGM is very sensitive to the timing and heat injection of the Epoch of Reionization \cite{Onorbe2017,Onorbe2017b}, complex and poorly understood physical processes related to galaxy formation have often be thought to only play a relatively minor role in determining its structure \cite{Kollmeier2006, Desjacques2006}, and are being neglected in most existing studies. However, as the statistical errors of DESI will significantly decrease compared to previous surveys, physical processes like star formation and stellar and Active Galactic Nuclei (AGN) feedback, which inject energy in the intergalactic medium, should be taken into consideration \cite{Viel2013,Chabanier2020, MonteroCamacho2021}. However, modeling all those processes from ``first principles'' and jointly exploring both the cosmological and astrophysical/nuisance parameters arising from these models will require computational capabilities that are not feasible in the near future, even using the largest high-performance computing systems.  We therefore foresee including the impact of galaxy formation and reionization on Ly-$\alpha$ statistics at the post-processing stage with analytical recipes established via auxiliary hydrodynamical simulations \cite{Chabanier2020}.

To summarize, for the Ly-$\alpha$ forest analyzes the following simulations are necessary:
\begin{itemize}
    \item Grids of $\sim 10$ hydrodynamical simulations with limited astrophysics (i.e. no feedbacks, no/simplified star formation, uniform reionization/heating) to model the dependencies of the overall IGM on cosmological parameters including the nature of dark matter, but providing enough resolution to resolve low-density IGM gas physics
    \item Additional full galaxy formation and reionization simulations as auxilliary inputs to derive analytical corrections for astrophysical effects
\end{itemize}


\section{Conclusion and Future Prospects}
The Milky Way and Ly-$\alpha$ constraints on the properties of dark matter described here require wide field, deep spectroscopy to create large samples of faint targets over a wide area of sky.  In the latter part of this decade, after the current DESI survey is finished, DESI will still have the highest survey speed of all the dedicated facilities available.  A next generation spectroscopic survey with DESI is being designed. The science goals include a high-density galaxy survey at $z<$1 to get improved precision on the dark energy equation of state in the cosmic acceleration epoch, an emission-line galaxy (ELG) survey at 1$<z<$1.6 for improved baryon acoustic oscillation (BAO) and red-shift distortion (RSD) studies, and a quasar and galaxy survey at $z>$2 that would give direct tracers for BAO and RSD as well as the possibility of mapping the Ly-$\alpha$ forest. The parameters of these surveys, which are designed to constrain the properties of dark energy, are well-matched to the dark matter investigations described in this document.  A next-generation DESI survey could include programs to map stellar velocities in the Milky Way halo and to acquire a larger quasar sample with higher magnitude that would improve the statistical precision of Ly-$\alpha$ cosmology constraints at DESI's resolution. Including these programs would allow the DESI facility to address both dark matter and dark energy science goals within the same next-generation survey program and operations plan. DOE funding for dark matter related science with DESI will expand the scope and impact of this unique and powerful observational facility.


\bibliographystyle{unsrt}
\bibliography{main.bib}

\end{document}

%% file: newcommands.tex
\newcommand{\kms}{\ensuremath{\,\mathrm{km\ s}^{-1}}}

\newcommand{\kmskpc}{\ensuremath{\,\mathrm{{km\ s}^{-1}\ {kpc}^{-1}}}}

\newcommand{\Gaia}{{\it Gaia}}

\newcommand{\msun}{\mbox{$\mathrm{M}_{\odot}$}}

\newcommand{\beq}{\begin{equation}}
\newcommand{\eeq}{\end{equation}}

%% file: authors.tex
\usepackage{authblk}

\author[1]{Monica Valluri}
\affil[1]{University of Michigan, Ann Arbor, MI 48109, USA}
\author[2]{Solene Chabanier}
\affil[2]{Lawrence Berkeley National Laboratory, Berkeley, CA 94720, USA}
\author[3,4]{Vid Ir\v{s}i\v{c}}
\affil[3]{Kavli Institute for Cosmology, University of Cambridge, Madingley Road, Cambridge CB3 0HA, UK}
\affil[4]{Cavendish Laboratory, University of Cambridge, 19 J. J. Thomson Ave., Cambridge CB3 0HE, UK}
\author[5]{Eric Armengaud}
\affil[5]{IRFU, CEA, Université Paris-Saclay, F-91191 Gif-sur-Yvette, France}
\author[6,7]{Michael Walther}
\affil[6]{University Observatory, Faculty of Physics, Ludwig-Maximilians-Universit\"at, Scheinerstr. 1, 81677 Munich, Germany}
\affil[7]{Excellence Cluster ORIGINS, Boltzmannstr. 2, 85748 Garching, Germany}
\author[8]{Connie Rockosi}
\affil[8]{University of California, Santa Cruz, CA 95064, USA}
\author[9]{Miguel A. S\'anchez-Conde}
\affil[9]{Instituto de F\'isica Te\'orica (IFT) UAM-CSIC and Departamento de F\'isica Te\'orica, Universidad Aut\'onoma de Madrid, Calle Nicol\'as Cabrera, 13-15, 28049 Madrid, Spain}
\author[1]{Leandro {Beraldo e Silva}}
\author[10]{Andrew P. Cooper}
\affil[10]{Department of Physics, National Tsing Hua University, Taiwan}
\author[11]{Elise Darragh-Ford}
\affil[11]{Kavli Institute for Particle Astrophysics and Cosmology and Department of Physics, Stanford University, CA 94305, USA; SLAC National Accelerator Laboratory, Menlo Park, CA 94025}
\author[12]{Kyle Dawson}
\affil[12]{University of Utah, Department of Physics and Astronomy, 115 S 1400 E, Salt Lake City, UT 84112, USA}
\author[13]{Alis J. Deason}
\affil[13]{Institute for Computational Cosmology, Durham University, Durham, DH1 3LE, UK}
\author[2]{Simone Ferraro}
\author[14]{Jaime E. Forero-Romero}
\affil[14]{Departamento de Física, Universidad de los Andes, Cra. 1 No. 18A-10 Edificio Ip, CP 111711, Bogotá, Colombia}
\author[15]{Antonella Garzilli}
\affil[15]{EPFL Laboratoire d’astrophysique, BSP 320 – Cubotron, CH – 1015 Lausanne, Switzerland}
\author[16]{Ting Li}
\affil[16]{David A. Dunlap, Dept of Astronomy and Astrophysics, University of Toronto, Ontario, M5S 3H4 Canada}
\author[2]{Zarija Luki\'c}
\author[17]{Christopher J. Manser}
\affil[17]{Astrophysics Group, Department of Physics, Imperial College London, Prince Consort Rd, London, SW7 2AZ, UK}
\author[2,5]{Nathalie Palanque-Delabrouille}
\author[5]{Corentin Ravoux}
\author[18]{Ting Tan}
\affil[18]{Sorbonne Universit\'e, CNRS/IN2P3, Laboratoire de Physique Nucl\'eaire et de Hautes Energies, LPNHE, 4 Place Jussieu, F-75252 Paris, France}
\author[19]{Wenting Wang}
\affil[19]{Shanghai Jiaotong University, Shanghai, China}
\author[11]{Risa H. Wechsler}
\author[13]{Andreia Carrillo}
\author[20]{Arjun Dey}
\affil[20]{NSF’s National Optical-Infrared Astronomy Research Laboratory, 950 N. Cherry Ave., Tucson, AZ 85719}
\author[21]{Sergey E. Koposov}
\affil[21]{Institute for Astronomy, University of Edinburgh, Royal Observatory, Blackford Hill, Edinburgh EH9 3HJ, UK}
\author[22]{Yao-Yuan Mao}
\affil[22]{Department of Physics and Astronomy, Rutgers, The State University of New Jersey, Piscataway, NJ 08854, USA}
\author[23]{Paulo Montero-Camacho}
\affil[23]{Department of Astronomy, Tsinghua University, China}
\author[24,25]{Ekta Patel}
\affil[24]{Department of Astronomy, University of California, Berkeley CA 94720, USA}
\affil[25]{Miller Institute for Basic Research in Science, University of California, Berkeley CA 94720, USA}
\author[26]{Graziano Rossi}
\affil[26]{Sejong University, Department of Physics and Astronomy, Seoul, 143-747, Republic of Korea}
\author[27]{L. Arturo Ureña-López}
\affil[27]{Universidad de Guanajuato, México}
\author[28]{Octavio Valenzuela}
\affil[28]{Universidad Nacional Autónoma de México, Instituto de Astronomía}